\documentclass{elsart}
\usepackage{graphicx}
\usepackage{amssymb}

\begin{document}

\begin{frontmatter}



\title{Universal quantum logic gates in a scalable Ising spin quantum computer}


\author{G. F. Mkrtchian}

\address{Department of Quantum Electronics, Yerevan State University,\\
1 A. Manukian,Yerevan 375025, Armenia}
\ead{mkrtchian@ysu.am}
\begin{abstract}
We consider the model of quantum computer, which is represented as a Ising
spin lattice, where qubits (spin-half systems) are separated by the
isolators (two spin-half systems). In the idle mode or at the single bit
operations the total spin of isolators is 0. There are no need of
complicated protocols for correcting the phase and probability errors due to
permanent interaction between the qubits. We present protocols for
implementation of universal quantum gates with the rectangular
radio-frequency pulses.
\end{abstract}

\end{frontmatter}


\section{Introduction}

The growing interest in quantum computation stimulates the search for new
schemes to prepare and manipulate qubits. Numerous proposals have been
introduced for experimental realization of a quantum computer (QC) \cite%
{Zel,N}. From the theoretical point of view QC could in principle be
realized by a one-dimensional array of simple two state systems, such as
single electron spins, coupled via the Heisenberg interaction (see, for
example, Refs. \cite{H1,H2,H3}). The Ising-spin chain has also been proposed
as a model system which allows to implement a quantum computer \cite%
{Lloyd,Berman}. Moreover, the switchable Heisenberg interaction alone can
provide the universal quantum computation \cite{H-alone}. However, in many
proposals for quantum computation, two bit operations, whose implementation
depends on the interaction between qubits, require to switch on and off the
couplings between qubits, which is not experimentally easy to realize. For
these schemes it is required precise control of the magnitude of the
Heisenberg interaction and effectively turn on and off it.

There are also proposals with the permanent interaction between qubits (see,
for example Refs. \cite{B2,B1}). However, perpetual untunable coupling of
qubits causes certain problems for realization of quantum gates, depending
on the particular form of the interaction. Particularly for the one
dimensional Ising spin chain QC complicated protocols with many \textit{rf}
pulses (for correcting phase and probability errors) are needed for single
and two qubit gates \cite{B1}, which requires high precision and complicates
the operation. For example, $5$ \textit{rf} pulses with the particular set
of phases are necessary for Not gate and $18$ pulses for Control-Not gate
between neighboring qubits.

The possibilities avoiding switching of interaction between qubits have been
investigated in Refs. \cite{Zh,Ben}. In Ref. \cite{Zh} it has been
considered architecture of the QC showing that the Heisenberg interaction
can be effectively negated by inserting EPR spin pairs (in singlet state)
between the information carrying qubits. In Ref. \cite{Zh} it has also been
considered architecture of the QC with Ising interaction. With the encoded
qubits (two physical qubits) Ising interaction can also be negated. In Ref. 
\cite{Ben} it has been shown that one can perform quantum computation in a
one dimensional Heisenberg chain in which the interactions are always on,
provided that one can abruptly tune the Zeeman energies of the individual
qubits.

Other schemes avoiding problems of perpetual coupling exist too. For
instance, in Ref. \cite{YB} authors describe an architecture of QC based on
a processing core where multiple qubits interact perpetually, and a separate
`store' where qubits exist in isolation. The obstacles connected with
implementation of single and two bit gates with permanent interaction
between qubits are also overcome in the proposal of QC, which is known as
\textquotedblleft one way quantum computer\textquotedblright\ \cite{One-way}%
. Here one can implement a quantum computer on a lattice of qubits with only
single bit measurements on a cluster states \cite{claster}, which are
created with the controllable Ising interaction.

In this work we consider the model of quantum computer, which is represented
as a Ising spin lattice where qubits (spin-half systems) are separated by
the isolators (two spin-half systems). This architecture is analogous to one
considered in Ref. \cite{Zh} for off diagonal interactions between qubits
where it was proposed to implement a virtual switch by carrying out the
steps of the quantum computation in and out of designed \textquotedblleft
interaction free subspaces\textquotedblright . In contrast to Ref. \cite{Zh}
the architecture considered in the present paper is based on the Ising type
interaction and it is not required turning on and off couplings between
isolators. Hence, this scheme allows to compute without switching the
couplings and simplifies the operation of QC compared to one dimensional
Ising spin chain QC \cite{B1}.

The paper is organized as follows. The Ising spin quantum computer model is
described in Sec.~II. The realization of the universal gates are considered
in Sec.~III. In Sec. IV we summarize our results.

\section{Ising spin quantum computer}

Architecture of the quantum computer with the Ising spin lattice is shown in
Fig. 1. It is assumed that information carrying qubits are situated in the
middle chain and do not directly interact. Qubits in the upper and lower
chains play the rule of isolators. We assume that in the idle mode or at the
single bit operations the spins of isolators are oppositely directed. In
this case for two neighboring qubits of middle chain, the total Ising
interaction is vanished for arbitrary states of qubits. This can be achieved
at the initialization of the system by two ways. Firstly, if one applies a
strong global field in the $z$ direction to all qubits, at low temperatures
all spins will line up with the field. Then one should flip the states of
the qubits situated at either above or below chains. Secondly, one can
assume that the global fields are inverted for the upper and lower spin
chains. Hence, if Larmor frequencies of qubits are much greater than Ising
interaction constant, then ground state of our model will coincide with the
desired initial state of QC. Then for the initialization of the QC one
should just cool the system. Note, that this architecture is also relevant
to a system with the Heisenberg interaction. As is known, if the frequency
difference between the neighboring spins is much greater than the spin-spin
interaction, then the Heisenberg interaction tends to an effective Ising
form \cite{H-Ising}. In this case the main technical challenge is the
creation of a large gradient of the magnetic field and the decoherence
caused by the sources of the magnetic field. The latest development of the
micropattern wires technique provides the magnetic field gradients $10^{6}\ 
\mathrm{T/m}$ \cite{Mag1,Mag2} and opens new possibilities for realization
of Ising spin QC.

The Hamiltonian for the Ising spin lattice (Fig. 1) placed in an external
nonuniform magnetic field can be represented as%
\[
\widehat{H}_{0}=-\sum_{k=0}^{N-1}\omega _{k}S_{k}^{z}+\sum_{k=0}^{N-2}\omega
_{ak}S_{ak}^{z}-\sum_{k=0}^{N-2}\omega _{bk}S_{bk}^{z} 
\]%
\begin{equation}
-2J_{0}\sum_{k=0}^{N-2}S_{ak}^{z}S_{bk}^{z}-2J\sum_{k=0}^{N-2}\left(
S_{k}^{z}+S_{k+1}^{z}\right) \left( S_{ak}^{z}+S_{bk}^{z}\right) .
\label{Ham}
\end{equation}

\vspace{0.1cm}Here $\hbar =1$, $S_{k}^{z}$ is the operator of the $z$
component of $k$th spin ($1/2$) with Larmor frequency $\omega _{k}$, $J$ is
the interaction constant between qubits and isolators, and $J_{0}$ is the
interaction constant between isolators. To distinguish the isolators from
the information carrying qubits we additionally label these qubits via
subscripts \textquotedblleft $a$\textquotedblright\ (upper qubits) and
\textquotedblleft $b$\textquotedblright\ (lower qubits).

The initial wave function of the QC can be represented as%
\begin{equation}
|\Psi _{0}\rangle =\bigotimes_{k=0}^{N-1}|0\rangle
_{k}\bigotimes_{k=0}^{N-2}|1_{a}\rangle _{k}|0_{b}\rangle _{k},  \label{entw}
\end{equation}%
with the energy%
\begin{equation}
\mathcal{E}=-\frac{1}{2}\left[ \sum_{k=0}^{N-1}\omega
_{k}+\sum_{k=0}^{N-2}\left( \omega _{ak}+\omega _{bk}\right) -J_{0}\left(
N-1\right) \right] .  \label{Energy}
\end{equation}

We assume that single bit gates are easy and fast to implement either by
individual addressing of single qubits or by selective electromagnetic
pulses whose frequencies are close (resonant transition) to the Larmor
frequencies of the desired qubits. For the latter case it has been assumed
nonuniform magnetic field in the Hamiltonian (\ref{Ham}). Since qubits are
not coupled to each other, fully parallel operations are possible. In
contrast to one dimensional Ising chain \cite{B1} in the considering model
there is no need of complicated protocols for correcting the phase and
probability errors due to permanent interaction between the qubits. As we
will see below two qubit universal gate between neighboring qubits can be
realized simply operating on the isolators.

\section{\protect\vspace{0.1cm}Universal Phase Gate}

In this section we describe the protocols for the two qubit universal gates.
For this purpose it is sufficient to consider one elementary cell, with two
qubits and isolators. The Hamiltonian of elementary cell in the external 
\textit{rf} field can be represented as%
\[
\widehat{H}_{cell}=-\omega _{1}S_{1}^{z}-\omega _{2}S_{2}^{z}+\omega
_{a1}S_{a1}^{z}-\omega _{b1}S_{b1}^{z}-2J\left( S_{1}^{z}+S_{2}^{z}\right)
\left( S_{a1}^{z}+S_{b1}^{z}\right) 
\]%
\begin{equation}
-2J_{0}S_{a1}^{z}S_{b1}^{z}-{\frac{\Omega }{2}}\left\{ S_{a}^{-}\exp \left[
-i\omega _{a}t\right] +S_{a}^{+}\exp \left[ i\omega _{a}t\right] \right\} ,
\label{HCEEL}
\end{equation}%
where $S^{\pm }=S^{x}\pm iS^{y}$, $\omega _{a}$ is the frequency of the
pulse, and $\Omega $ is the Rabi frequency.

We assume that the initial state is%
\[
|\Psi _{cell}\rangle =|1_{a}\rangle |0_{b}\rangle \left[ C_{0}\left(
0\right) |0\rangle |0\rangle +C_{1}\left( 0\right) |0\rangle |1\rangle
\right. 
\]%
\begin{equation}
\left. +C_{2}\left( 0\right) |1\rangle |0\rangle +C_{3}\left( 0\right)
|1\rangle |1\rangle \right]  \label{iwf}
\end{equation}%
and our goal is to realize the controlled phase gate:

\begin{figure}[tbp]
\includegraphics[]{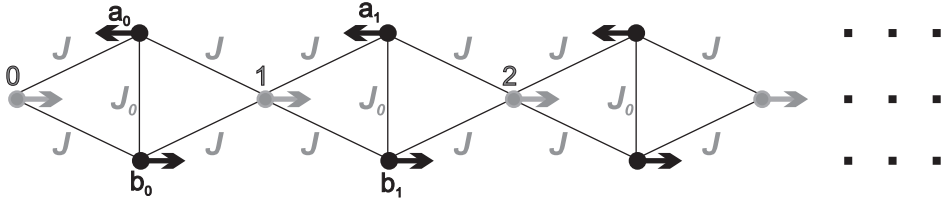}
\caption{Architecture of the quantum computer with the Ising spin lattice.
The information carrying qubits are midle line qubits. Qubits situated at
above and below chains are \textquotedblleft isolators\textquotedblright .
It is assumed that information carrying qubits do not directly interact
(situated relatively far from each other). In the idle mode or at the single
bit operations the spins of isolators are oppositely directed.}
\label{eps1.1}
\end{figure}
\[
CPHASE|\Psi _{cell}\rangle =e^{-i\Theta }|1_{a}\rangle |0_{b}\rangle \left[
C_{0}\left( 0\right) |0\rangle |0\rangle +C_{1}\left( 0\right) |0\rangle
|1\rangle \right. 
\]%
\begin{equation}
\left. +C_{2}\left( 0\right) |1\rangle |0\rangle -C_{3}\left( 0\right)
|1\rangle |1\rangle \right] ,  \label{cphase}
\end{equation}%
where $\Theta $ is the possible overall phase.

Under the condition $\omega _{a}\simeq \omega _{a1}$ ($\Omega ,J\ll \omega
_{a}$) the pulse effectively affects only above located spin (resonant
transition). In this case, evolution of the wave function (\ref{iwf}) is
confined within subspace (8 dimensional) of the entire Hilbert space of four
qubits and can be written in the form (in the interaction picture)%
\[
|\Psi \left( t\right) \rangle =|1_{a}\rangle |0_{b}\rangle \left[
C_{0}\left( t\right) e^{-i\mathcal{E}_{0}t}|0\rangle |0\rangle +C_{1}\left(
t\right) e^{-i\mathcal{E}_{1}t}|0\rangle |1\rangle \right. 
\]%
\[
\left. +C_{2}\left( t\right) e^{-i\mathcal{E}_{2}t}|1\rangle |0\rangle
+C_{3}\left( t\right) e^{-i\mathcal{E}_{3}t}|1\rangle |1\rangle \right] 
\]%
\[
+|0_{a}\rangle |0_{b}\rangle \left[ \widetilde{C}_{0}\left( t\right) e^{-i%
\widetilde{\mathcal{E}}_{0}t}|0\rangle |0\rangle +\widetilde{C}_{1}\left(
t\right) e^{-i\widetilde{\mathcal{E}}_{1}t}|0\rangle |1\rangle \right. 
\]%
\begin{equation}
\left. +\widetilde{C}_{2}\left( t\right) e^{-i\widetilde{\mathcal{E}}%
_{2}t}|1\rangle |0\rangle +\widetilde{C}_{3}\left( t\right) e^{-i\widetilde{%
\mathcal{E}}_{3}t}|1\rangle |1\rangle \right] ,  \label{wf}
\end{equation}%
where $C_{j}(t)$ and $\widetilde{C}_{j}(t)$ ( $j=0,1,2,3$) are probability
amplitudes of the states with energies:%
\[
\mathcal{E}_{0}=-\frac{\omega _{1}+\omega _{2}}{2}+\mathcal{E}_{ab};\quad 
\widetilde{\mathcal{E}}_{0}=\mathcal{E}_{0}+\omega _{a1}-J_{0}-2J, 
\]%
\[
\mathcal{E}_{1}=\frac{\omega _{2}-\omega _{1}}{2}+\mathcal{E}_{ab};\quad 
\widetilde{\mathcal{E}}_{1}=\mathcal{E}_{1}+\omega _{a1}-J_{0}, 
\]%
\[
\mathcal{E}_{2}=\frac{\omega _{1}-\omega _{2}}{2}+\mathcal{E}_{ab};\quad 
\widetilde{\mathcal{E}}_{2}=\mathcal{E}_{2}+\omega _{a1}-J_{0}, 
\]%
\[
\mathcal{E}_{3}=\frac{\omega _{1}+\omega _{2}}{2}+\mathcal{E}_{ab};\quad 
\widetilde{\mathcal{E}}_{3}=\mathcal{E}_{3}+\omega _{a1}-J_{0}+2J, 
\]%
\begin{equation}
\mathcal{E}_{ab}=-\frac{\omega _{b1}+\omega _{a1}-J_{0}}{2}.  \label{ens}
\end{equation}%
From the Schr\"{o}dinger equation%
\begin{equation}
i\frac{\partial |\Psi \left( t\right) \rangle }{\partial t}=\widehat{H}%
_{cell}|\Psi \left( t\right) \rangle  \label{SE}
\end{equation}%
one can obtain the equations for the probability amplitudes $C_{j}(t)$ and $%
\widetilde{C}_{j}(t)$. System of coupled differential equations for the
coefficients $C_{j}(t)$, $\widetilde{C}_{j}(t)$ splits into four independent
groups. Each group consists of two equations of the form%
\[
i\frac{dC_{j}(t)}{dt}=-\frac{\Omega }{2}e^{i\Delta _{j}t}\widetilde{C}%
_{j}(t), 
\]%
\begin{equation}
i\frac{d\widetilde{C}_{j}(t)}{dt}=-\frac{\Omega }{2}e^{-i\Delta
_{j}t}C_{j}(t),\ j=0,1,2,3.  \label{eqc}
\end{equation}%
The detunings in Eq. (\ref{eqc}) are 
\[
\Delta _{1}=\Delta _{2}\equiv \Delta =\omega _{a}-\omega _{a1}+J_{0}, 
\]%
\begin{equation}
\Delta _{0}=\Delta +2J,\quad \Delta _{3}=\Delta -2J.  \label{detun}
\end{equation}%
Assuming rectangular \textit{rf} pulse, for the initial conditions%
\begin{equation}
C_{j}\left( t=0\right) =C_{j}\left( 0\right) ,\qquad \widetilde{C}_{j}\left(
t=0\right) =0,  \label{incond}
\end{equation}%
the solution of Eq. (\ref{eqc}) is%
\[
C_{j}\left( \tau \right) =C_{j}\left( 0\right) e^{i\frac{\Delta _{j}}{2}\tau
}\left( \cos \frac{\lambda _{j}}{2}\tau -\frac{i\Delta _{j}}{\lambda _{j}}%
\sin \frac{\lambda _{j}}{2}\tau \right) , 
\]%
\begin{equation}
\widetilde{C}_{j}\left( \tau \right) =C_{j}\left( 0\right) \frac{i\Omega }{%
\lambda _{j}}e^{-i\frac{\Delta _{j}}{2}\tau }\sin \frac{\lambda _{j}}{2}\tau
,  \label{sol}
\end{equation}%
where $\tau $ is the pulse duration and $\lambda _{j}=\sqrt{\Delta
_{j}^{2}+\Omega ^{2}}$.

For the realization of controlled phase gate (\ref{cphase}) we will consider
two cases, depending on the magnitude of spin-spin interaction. If the
spin-spin interaction is strong enough, i.e. $J>>$ $\Omega $, then under the
resonance condition $\Delta =2J$ we have%
\begin{equation}
\Delta _{1}=\Delta _{2}\equiv \Delta =2J,\ \Delta _{0}=4J,\ \Delta _{3}=0
\label{rd}
\end{equation}%
and for the couplings%
\[
\lambda _{0}=\sqrt{16J^{2}+\Omega ^{2}}\simeq 4J, 
\]%
\begin{equation}
\ \lambda _{1,2}=\sqrt{4J^{2}+\Omega ^{2}}\simeq 2J,\ \lambda _{3}=\Omega .
\label{rc}
\end{equation}%
The states $|0_{a}\rangle |0_{b}\rangle |0\rangle |0\rangle $, $%
|0_{a}\rangle |0_{b}\rangle |0\rangle |1\rangle $, and $|0_{a}\rangle
|0_{b}\rangle |1\rangle |0\rangle $ will be shifted out of resonance. Their
probabilities are small and proportional to $\Omega ^{2}/J^{2}<<1$ and may
be neglected:%
\begin{equation}
C_{j}\left( \tau \right) =C_{j}\left( 0\right) ,\ \widetilde{C}_{j}\left(
\tau \right) \simeq 0,\ j\neq 3.  \label{nonrs}
\end{equation}%
For the resonant states we have%
\begin{equation}
C_{3}\left( \tau \right) =C_{3}\left( 0\right) \cos \frac{\Omega }{2}\tau ,\ 
\widetilde{C}_{3}\left( \tau \right) =iC_{3}\left( 0\right) \sin \frac{%
\Omega }{2}\tau ,  \label{resonants}
\end{equation}%
and the $2\pi $ pulse with the pulse duration $\tau =2\pi /\Omega $ will be
sufficient for the CPHASE gate (\ref{cphase}) (at that isolator returns to
its initial state). In this case the clock speed of QC will be $\sim \Omega $%
.

If $\Omega \sim J$, then all states will be excited and as it follows from
Eq. (\ref{sol}) for the Control-Phase gate one should require%
\[
\frac{\lambda _{0}}{2}\tau =2\pi k_{0},\quad \frac{\lambda _{1,2}}{2}\tau
=2\pi k_{1}
\]%
\begin{equation}
\frac{\lambda _{3}}{2}\tau =\pi \left( 1+2k_{3}\right) ;\
k_{0,1,3}=1,2,3,....  \label{mcond}
\end{equation}%
Solving these equations for the couplings and detuning we obtain%
\[
\lambda _{0}^{2}=\frac{32k_{0}^{2}J^{2}}{4k_{0}^{2}-8k_{1}^{2}+\left(
1+2k_{3}\right) ^{2}},\quad \lambda _{1,2}^{2}=\frac{32k_{1}^{2}J^{2}}{%
4k_{0}^{2}-8k_{1}^{2}+\left( 1+2k_{3}\right) ^{2}},
\]%
\begin{equation}
\lambda _{3}^{2}=\frac{8\left( 1+2k_{3}\right) ^{2}J^{2}}{%
4k_{0}^{2}-8k_{1}^{2}+\left( 1+2k_{3}\right) ^{2}},\quad \Delta =J\frac{%
4k_{0}^{2}-\left( 1+2k_{3}\right) ^{2}}{4k_{0}^{2}-8k_{1}^{2}+\left(
1+2k_{3}\right) ^{2}}.  \label{nonr}
\end{equation}%
Particularly for $k_{0}=k_{1}=k_{3}=1$ from Eqs. (\ref{nonr}) we have%
\begin{equation}
\Delta =-J,\ \Delta _{0}=J,\ \Delta _{3}=-3J,\ \Omega =\sqrt{\frac{27}{5}}J,
\label{par}
\end{equation}%
and for the pulse duration from Eq. (\ref{mcond}) we obtain $\tau =\sqrt{5/2}%
\pi /J$. The resulting unitary transformation is%
\begin{equation}
C_{\Phi }=e^{-i\frac{J}{2}\tau }\left( 
\begin{array}{cccc}
e^{iJ\tau } & 0 & 0 & 0 \\ 
0 & 1 & 0 & 0 \\ 
0 & 0 & 1 & 0 \\ 
0 & 0 & 0 & -e^{-iJ\tau }%
\end{array}%
\right) ,  \label{CPH}
\end{equation}%
which is an entangling gate. In this case the clock speed of QC will be $%
\sim J$. It is easy to show that this gate, along with two suitable
single-qubit gates (rotations around the z axis), generates the known
universal phase gate: 
\begin{equation}
CPHASE=R_{z}^{(1)}\left( -\frac{J}{2}\tau \right) R_{z}^{(2)}\left( -\frac{J%
}{2}\tau \right) C_{\Phi }=e^{i\Theta }\left( 
\begin{array}{cccc}
1 & 0 & 0 & 0 \\ 
0 & 1 & 0 & 0 \\ 
0 & 0 & 1 & 0 \\ 
0 & 0 & 0 & -1%
\end{array}%
\right) ,  \label{TCPH}
\end{equation}%
where the overall phase is $\Theta =-\pi \sqrt{5/8}$. It is simple to use
established formalisms \cite{Zel,N} to generate a CNOT gate by sandwiching a
CPHASE gate between $1$ and $2$ qubits with Hadamard gate $H_{2}$.

\section{Conclusion}

In conclusion, we have described the model of quantum computer, which is
represented as a Ising spin lattice where qubits are separated by the
isolators. For this architecture single and two qubit gates are easy and
fast to implement either by individual addressing of single spins or by
selective electromagnetic pulse in contrast to one dimensional Ising chain
where due to permanent interaction between qubits complicated protocols for
correcting the phase and probability errors are needed \cite{B1}. We
discussed two possible physical realizations of the universal phase gate;
one when Rabi frequency is much smaller than Ising interaction constant and
the other when Rabi frequency is larger or comparable with Ising interaction
constant. In both cases the universal phase gate can be realized with the
single pulse applied to isolator. Compared with the similar proposal of QC
with encoded qubits \cite{Zh}, considered architecture is less complex in
terms of the steps involved in generating one and two qubit gates.

We gratefully acknowledge helpful discussions with Prof. H. K. Avetissian.

\end{document}